\documentclass[runningheads,citeauthoryear]{apinv}
\usepackage{epsfig,cite,graphics}
\usepackage[T2A]{fontenc}
\usepackage[bulgarian,english]{babel}

\begin{document}
\setcounter{page}{26}
\title{Photometric reverberation mapping\\ of Markarian~279}
\titlerunning{Mapping of Mrk 279}
\author{R.~Bachev, A.~Strigachev, E.~Semkov, S.~Boeva, S.~Peneva, S.~Ibryamov, K.~Stoyanov, B.~Spassov, S.~Tsvetkova, B.~Mihov, G.~Latev, D.~Dimitrov}
\authorrunning{Bachev et al.}
\tocauthor{R.~Bachev, A.~Strigachev, E.~Semkov, S.~Boeva, S.~Peneva, S.~Ibryamov, K.~Stoyanov, B.~Spassov, S.~Tsvetkova, B.~Mihov, G.~Latev, D.~Dimitrov}
\institute{Institute of Astronomy and NAO, Bulgarian Academy of Sciences, BG-1784 Sofia
      \email{bachevr@astro.bas.bg} }
\papertype{Submitted on 10.10.2013; Accepted on 26.10.2013.}
\maketitle
\begin{abstract}
By using standard broad-band VRI photometry we were able to discriminate the variations of the broad hydrogen 
alpha line from the continuum variations for the active galaxy Mkn~279. Cross-correlating both light curves 
enabled us to determine the time lag of the broad line variations behind the continuum and thus to determine 
the BLR size (about 8 light days). Our preliminary results are rather consistent with the spectroscopic 
reverberation mapping results (about 12/17 days). This study is a part of an ambitious program to perform 
photometric reverberation mapping and determine BLR sizes (respectively -- the central black hole masses) 
for more that 100 nearby AGN.
\end{abstract}



\section*{1. Introduction}

Almost the only way to measure directly the super-massive black hole (SMBH) in the centers of active galactic nuclei (AGN) 
is through a technique, called reverberation mapping (a review by Peterson \& Horne, 2004). This technique is based on 
the geometrical path difference between the ionizing continuum and the broad line clouds, where the broad emission 
presumably originates, thus leading to a certain time delay between the continuum and the line variations. Measuring 
this delay provides an accurate estimate of the linear distance to the broad line region (BLR). Thus, knowing the 
width of the lines (Keplerian motion of the BLR clouds is assumed) and the distance, one can calculate the mass 
inside the BLR, which is significantly dominated by the SMBH. 

Although very promising, reverberation mapping technique has been applied so far for only about 50 AGN (Bentz et al. 2010). 
The main obstacle here is that the spectroscopic monitoring (neede to measure the emission line fluxes) requires typically a 2-m class 
telescope. A number of reasons like restricted telescope time, etc., make projects like this difficult to organize and perform, 
especially what concerns the need of a dense monitoring. Fortunately, very recently it has been developed a modification to this 
technique, called \textit{\textbf{photometric}} reverberation mapping (Haas et al., 2011; Chelouche \& Daniel, 2012) which requires much smaller 
telescopes. The idea behind this novel approach is to use narrow-band filters, cleverly centered on the emission lines and 
the ambient continuum, instead of performing spectroscopic monitoring. Since the line response time is much longer 
than the continuum response time, one can successfully use the nearby optical continuum bands for a proxy for the central 
UV/X-ray changes. For the brightest objects, photometric reverberation mapping can successfully be performed with as small 
as 15-cm telescope (Haas et al., 2011).

Recently, photometric reverberation mapping has been applied to a number of objects 
(Edri et al., 2012; Pozo Nu\~{n}ez et al., 2012; Pozo Nu\~{n}ez et al., 2013;  Carroll åò àë., 2013) with results very close to the spectroscopic ones. 

In this work we report preliminary results of the photometric reverberation mapping technique applied 
to Seyfert~1 AGN Markarian~279. 

\section*{2. Observations}

We monitor photometrically on a regular basis Mkn~279 for the last $\sim$15 years using the 60-cm telescope of 
\textit{Belogradchik Observatory}; 200-cm, 50/70-cm, and 60-cm telescopes of \textit{Rozhen National Observatory} 
(Bulgaria); as well as 130-cm \textit{Skinakas} telescope (Greece). All telescopes are equipped with CCD and $BVRI$ 
filter sets. Initially the object was in a rather bright state (years 1998 -- 2007), but later faded significantly 
and remained in a low state during the years 2008 -- 2012. $BVRI$ light curves of Mkn~279 are shown in Fig.~1. 
A total of about $\sim$240 photometric points with typical photometric errors between $\sim$0.01 (VRI bands) 
and $\sim$0.03 ($B$-band) were collected during the entire observational period. 
A portion of these data was published in a previous study (Bachev \& Strigachev, 2004).

\begin{figure}[!htb]
  \begin{center}
    \centering{\epsfig{file=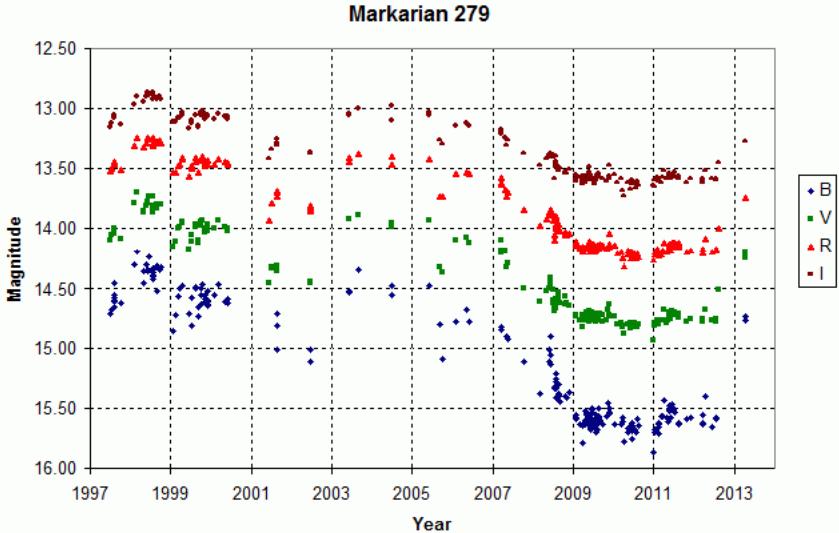, width=0.9\textwidth}}
\caption[]{$BVRI$ light curves of Mkn 279. All magnitudes are measured in an 8$\arcsec$ aperture with 
no correction made for the host galaxy contribution and Galaxy extinction}
    \label{f1}
  \end{center}
\end{figure}

\section*{3. Photometric reverberation mapping}

Photometric reverberation mapping proved to be a successful technique for determining BLR linear sizes, especially if 
narrow-band filters, centered on the lines of interest are used. As we traditionally monitored Mkn~279 with broad-band 
$BVRI$ filters, here we apply this technique using data, collected in these bands (so called \textit{\textbf{broadband}}
photometric mapping; Edri et al., 2012). Thus, the $R$-band flux can be considered as sum of the broad $H_{\alpha}$ and the 
continuum contributions, while $V$ and $I$ fluxes are proxies for the continuum (the constant narrow lines and the 
host galaxy do not affect this analysis). Fig.~2 shows the $VRI$ filter transmittances superimposed on the Mkn~279 
optical spectrum. $B$-band data is not used in this analysis due to generally larger photometric errors.

\begin{figure}[!htb]
  \begin{center}
    \centering{\epsfig{file=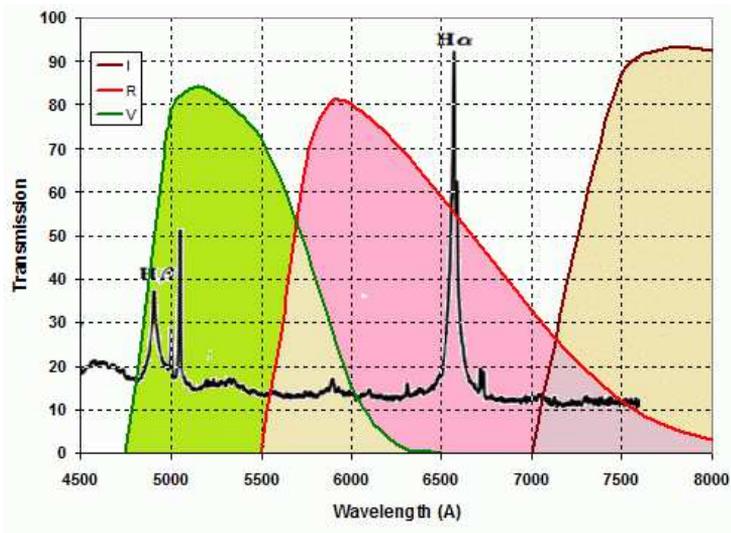, width=0.75\textwidth}}
\caption[]{$VRI$ filter transmittances superimposed on the Mkn~279 optical spectrum. Note that broad $H_{\beta}$ slightly 
affects the $V$-band}
    \label{f2}
  \end{center}
\end{figure}

In order to find the time delay of the broad lines ($H_{\alpha}$ in this case), we therefore cross-correlate the continuum 
(represented as an average of $V$ and $I$ band fluxes in order to minimize the errors) and the $R$-band flux, subtracted by 
the already calculated continuum (appropriately scaled), which thus represents the pure $H_{\alpha}$ emission. In other words, we 
search for the maximum of the following function:

$CCF(\tau) = \int_{-\infty}^{+\infty}(F_{R}(t+\tau) - F_{VI}(t+\tau)).F_{VI}(t)dt$,

\hspace{-6mm}where $F_{R}$ and $F_{VI}$ are fluxes, based on the measured $VRI$ magnitudes.

One can note that the broad $H_{\beta}$ slightly enters the $V$-band, which, in principle may compromise the continuum 
measurement. However, due to the low transmittance of $V$-band filter there, as well as the fact that $H_{\beta}$ influence 
is much weaker compared to $H_{\alpha}$ our understanding is that the effect is negligible. 

\section*{4. Results}

For this analysis we used only the first $\sim$70 observational points (years 1997.5 -- 2000.5) since during this period the 
object was bright (low photometric errors), active (significantly variable on short time scales) and the monitoring 
cadence was good enough. To find the cross-correlation between unevenly spaced datasets, we used the interpolation 
cross-correlation method (Gaskel \& Sparke, 1986), where the missing parts were linearly interpolated. We found a clear 
maximum (Fig.~3) for $\tau \approx$ 8 days. These first results are very encouraging as they are close to the spectroscopic 
reverberation values 12$\pm$3 days (Maoz et al., 1990) and 17$\pm$4 days (Bentz et al., 2009). A detailed analysis, including uncertainty 
estimates, will be published elsewhere (Bachev et al., to be published).

\begin{figure}[!htb]
  \begin{center}
    \centering{\epsfig{file=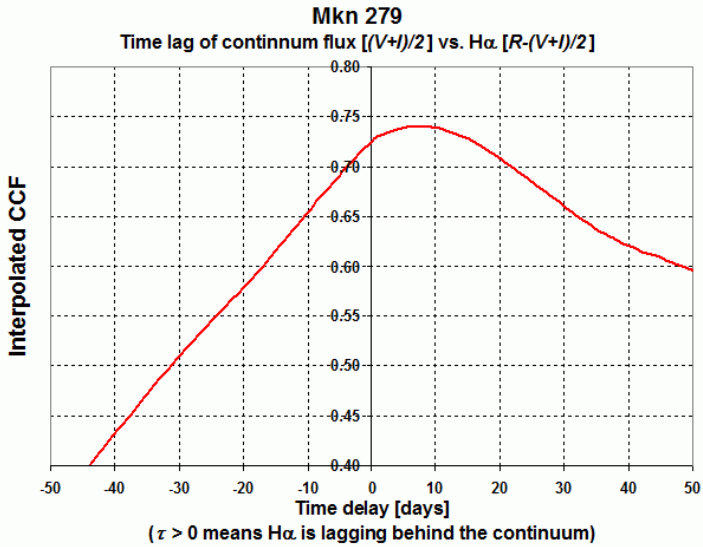, width=0.75\textwidth}}
\caption[]{Interpolated CCF, showing a delay of $H_{\alpha}$ behind the continuum for Mkn~279 of about 8~days}
    \label{f3}
  \end{center}
\end{figure}

Thus, knowing the width of the broad lines ($\nu$) as well as the linear distance to the BLR clouds (in this 
case $\sim$8 light days) and assuming virial motion, one can calculate the central black hole mass:

$M_{BH} \cong \frac{c}{G} \tau . \nu^{2}$,

which in our case gives $M_{BH}\approx3.10^{7}M_{\odot}$ (assuming $\nu\approx FWHM(H_{\alpha}) \approx$ 3050 km/s, our data, 
to be published elsewhere).

\section*{5. Conclusions}

Photometric reverberation mapping, even applied to broad-band filters, proved to be an effective, powerful, yet ``cheap''
technique for estimating the BLR size and the central black hole mass. In this work we applied this technique to the active 
galaxy nucleus Markarian~279, obtaining encouraging results, rather close to the spectroscopic mapping results. These 
results indicate that a dense enough photometric monitoring of AGN, even with relatively small instruments can be productive 
and scientifically significant.

{\it Acknowledgments:}This research was partially supported by Bulgarian NSF through grants DO 02-85 and DO 02-137 (2009). 
The Skinakas Observatory is a collaborative project of the University of Crete, the Foundation for Research and 
Technology -- Hellas, and the Max-Planck-Institut f\"ur Extraterrestrische Physik. We thank our referee Prof. Ts. Georgiev 
for his suggestions to improve this paper. 

{}

\end{document}